\documentclass[12pt]{article}
\usepackage{graphicx,amssymb,bbold,bm}
\usepackage{amsmath,amsfonts,epsfig}
\usepackage{braket}
\usepackage[usenames]{color} %%%%%%%%%%%
\usepackage{wrapfig}
\usepackage{amsmath,amsfonts,amssymb,bm}
\usepackage{amsfonts,bm}
\usepackage{tikz}
\usepackage{mathtools}

\usepackage{cancel}

\usepackage{ulem}

\newcommand{\p}{\partial}
\newcommand{\pslash}{p\kern-1ex /}
\newcommand{\qslash}{q\kern-1ex /}
\newcommand{\lslash}{l\kern-1ex /}
\newcommand{\sslash}{s\kern-1ex /}
\newcommand{\kaslash}{k_a\kern-2ex /}
\newcommand{\kbslash}{k_b\kern-2ex /}
\newcommand{\Dslash}{{\cal D}\kern-1.5ex /}

\newcommand{\bc}{\overline{c}}

\newcommand{\beqa}{\begin{eqnarray}}
\newcommand{\eeqa}{\end{eqnarray}}

\newcommand{\bpm}{\begin{pmatrix}}
\newcommand{\epm}{\end{pmatrix}}
\newcommand{\bbm}{\begin{bmatrix}}
\newcommand{\ebm}{\end{bmatrix}}

\def\p{\partial}

\newcommand{\Exp}[1]{\left\langle~#1~\right\rangle}
\makeatletter

\@addtoreset{equation}{section}
\makeatother
\usepackage{subfig}
\begin{document}

\setlength{\textheight}{230mm}      % テキストの高さ
\setlength{\headheight}{5mm}        % ヘッダの高さ
\setlength{\headsep}{0mm}           % テキストの最上部とヘッダの最下部との間隔
\setlength{\footskip}{10mm}         % テキストの最下部とフッタの最下部との間隔
\setlength{\topmargin}{30mm}        % 上のマージン
\addtolength{\topmargin}{-1in}      % 元の 1in の空白を削除

%\setlength{\unitlength}{15mm}

% \setlength{\topmargin}{-2.0truecm}
% \setlength{\textheight}{25.0truecm}
% \setlength{\textwidth}{16.0truecm}

% \voffset -0.7 true cm
% \hoffset 1.5 true cm
% \topmargin 0.0in
% %\evensidemargin 0.0in
% \oddsidemargin 0.0in
% \textheight 8.6in
% \textwidth 7.0in %5.4in
% \parskip 9 pt

\def\Tr{\hbox{Tr}}
\newcommand{\be}{\begin{align}}
\newcommand{\ee}{\end{align}}
\newcommand{\bea}{\begin{eqnarray}}
\newcommand{\eea}{\end{eqnarray}}
\newcommand{\beas}{\begin{eqnarray*}}
\newcommand{\eeas}{\end{eqnarray*}}
\newcommand{\nn}{\nonumber}
\newcommand{\bdyg}{\mathcal{G}}
\newcommand{\bdyR}{\mathcal{R}}
\newcommand{\tbdy}{\text{bdy}}
\newcommand{\bmcalT}{\bm{\calT}}
\newcommand{\pExp}[1]{\langle~#1~\rangle}
\newcommand{\calT}{\mathcal{T}}
\font\cmsss=cmss8
\def\C{{\hbox{\cmsss C}}}
\font\cmss=cmss10
\def\bigC{{\hbox{\cmss C}}}
\def\scriptlap{{\kern1pt\vbox{\hrule height 0.8pt\hbox{\vrule width 0.8pt
  \hskip2pt\vbox{\vskip 4pt}\hskip 2pt\vrule width 0.4pt}\hrule height 0.4pt}
  \kern1pt}}
\def\ba{{\bar{a}}}
\def\bb{{\bar{b}}}
\def\bc{{\bar{c}}}
\def\bphi{{\Phi}}
\def\Bigggl{\mathopen\Biggg}
\def\Bigggr{\mathclose\Biggg}
\def\Biggg#1{{\hbox{$\left#1\vbox to 25pt{}\right.\n@space$}}}
\def\n@space{\nulldelimiterspace=0pt \m@th}
\def\m@th{\mathsurround = 0pt}

\begin{titlepage}

\begin{center}

\vspace{5mm}
{\Large \bf {Symmetry breaking of $3$-dimensional AdS in holographic semiclassical gravity}} \\ [2pt]

\vspace{5mm}

\renewcommand\thefootnote{\mbox{$\fnsymbol{footnote}$}}
Akihiro Ishibashi${}^{1}$, 
Kengo Maeda${}^{2}$ and 
Takashi Okamura${}^{3}$

\vspace{2mm}

${}^{1}${\small \sl Department of Physics and Research Institute for Science and Technology, } \\ 
{\small \sl Kindai University, Higashi-Osaka, Osaka 577-8502, JAPAN}

${}^{2}${\small \sl Faculty of Engineering, Shibaura Institute of Technology,} \\   
{\small \sl Saitama 330-8570, JAPAN} 
%\\ 

${}^{3}${\small \sl Department of Physics and Astronomy, Kwansei Gakuin University,} \\   
{\small \sl Sanda, Hyogo, 669-1330, JAPAN} 
%\\ 

%\vspace{3mm}

{\small \tt 
{akihiro at phys.kindai.ac.jp},  {maeda302 at sic.shibaura-it.ac.jp} \\
{tokamura at kwansei.ac.jp}
}

\end{center}

%\vspace{2mm}

\noindent

\abstract{
We show that $3$-dimensional AdS spacetime can be semiclassically unstable due to strongly interacting quantum field effects. In our previous paper, we have pointed out the possibility of such an instability of AdS$_3$ by inspecting linear perturbations of the (covering space of) static BTZ black hole with AdS${}_4$ gravity dual in the context of holographic semiclassical problems. 
In the present paper, we further study this issue from thermodynamic viewpoint by constructing asymptotically AdS$_3$ semiclassical solutions 
% with vanishing source term (vacuum expectation value for stress-energy tensor) 
and computing free energies of the solutions. 
We find two asymptotically AdS${}_3$ solutions to the semiclassical Einstein equations with non-vanishing source term: the one whose free energy is smaller than that of the BTZ with vanishing source term and the other whose free energy is smaller than that of the global AdS$_3$ with no horizon (thus manifestly zero-temperature background). 
The instability found in this paper implies the breakdown of the maximal symmetries of AdS$_3$, and its origin is different from the well-known semiclassical linear instability since our holographic semiclassical Einstein equations in $3$-dimensions do not involve higher order derivative terms.     
} 
\end{titlepage}

\renewcommand\thefootnote{\mbox{\arabic{footnote}}}

%\tableofcontents
%\newpage
%%%%%%%%%%%%%%%%%%%%%
\section{Introduction}
%%%%%%%%%%%%%%%%%%%%%
One of the important issues in quantum general relativity is whether spacetime is stable under quantum effects. One approach to addressing such a problem is the semiclassical approximation in which gravitational field is treated classically, while matter fields quantum mechanically: Classical gravity obeys the semiclassical Einstein equations sourced by the vacuum expectation value of the renormalized stress-energy tensor for quantum matter fields. In this approach, Minkowski spacetime, for example, was found to be unstable against a certain type 
of quantum fluctuations~\cite{Horowitz:1978fq,Horowitz:1980fj,Suen:1989bg}. However, it is in general difficult to analyze such semiclassical problems for curved spacetimes, except for a few special cases~(see e.g.,~\cite{Starobinsky:1980te, Vilenkin:1985md}).    

Recently, the semiclassical Einstein equations have been reformulated in the holographic context~\cite{Compere:2008us,Ishibashi:2023luz}, in which $d$-dimensional metric on the conformal boundary of $(d+1)$-dimensional anti-de Sitter (AdS$_{d+1}$) bulk spacetime is promoted to be a dynamical field induced by boundary quantum conformal field theory (CFT). In this formulation, the $d$-dimensional semiclassical Einstein equations can be viewed as a mixed boundary conditions for the $(d+1)$-dimensional bulk classical metric, and the vacuum expectation value for quantum matter fields---whose evaluation is one of the hardest parts of the job in semiclassical problems---can be explicitly computed by exploiting the well-known formulas~\cite{deHaro:2000vlm} of the AdS/CFT correspondence~\cite{Maldacena:1997re, Gubser:1998bc, Witten:1998qj}. In this way, the holographic approach considerably simplifies the problem of how to set up the semiclassical Einstein equations, especially how to compute the source term, and at the same time makes it possible to analyze the effects of strongly coupled quantum fields on the dynamics of classical gravity.

In our previous paper \cite{Ishibashi:2023luz}, by taking advantage of the holographic approach mentioned above, we have analyzed the semiclassical Einstein equations and shown that the covering space of $3$-dimensional static BTZ black hole is semiclassically unstable under linear perturbations due to strongly coupled CFTs. 
We have introduced the universal parameter $\gamma_3$ which determines the onset of semiclassical instabilities. 
We have also shown the existence of a $3$-dimensional static asymptotically AdS semiclassical solution with a non-zero expectation value for stress-energy tensor, which may be interpreted as an asymptotically AdS black hole with ``quantum hair." 
  
In this paper, we further study the issue of holographic semiclassical instability of AdS$_3$ from the thermodynamic viewpoint. 
For this purpose, we investigate perturbations of semiclassical AdS$_3$ with vanishing expectation value of the stress-energy tensor and evaluate free energy by calculating the on-shell action at second order in perturbation. We find that the only non-zero terms in our action for the holographic setting with AdS$_4$ bulk and AdS$_3$ boundary are the $2$-dimensional surface terms of semiclassical AdS$_3$ solution. 
By adding appropriate counter term with respect to the $2$-dimensional surface, we find that the free energy for the semiclassical solution with non-vanishing stress energy tensor is always smaller than that of the background AdS$_3$ solution with vanishing expectation values for the stress-energy tensor. 
For comparison with our previous work~\cite{Ishibashi:2023luz}, we perform the analysis in both the covering space of static BTZ black hole (i.e., AdS$_3$ background with Killing horizon) as well as the global AdS$_3$ (i.e., the manifestly zero-temperature background with no horizon). For both cases, we arrive at the same conclusion that AdS$_3$ as a solution to the semiclassical Einstein equations with vanishing source term can be thermodynamically unstable, of which onset is determined by the control parameter $\gamma_3$ introduced in~\cite{Ishibashi:2023luz}. In particular, it is clear from the analysis of the global AdS$_3$ case that the quantum field on our AdS$_3$ is in the conformal vacuum state. This instability implies that the maximal symmetries of AdS$_3$ break down spontaneously, suggesting that a phase transition occurs between the semiclassical AdS$_3$ solution with vanishing stress-energy tensor and that with non-vanishing stress-energy tensor. This is a new instability, different from the well-known semiclassical linear instaiblity~\cite{Horowitz:1978fq,Horowitz:1980fj,Suen:1989bg,Simon:1990jn,Simon:1991bm} since our holographic semiclassical Einstein equations do not involve higher order derivative terms. 

This paper is organized as follows. In section~\ref{sec:2}, we will provide a general prescription for deriving 
the second variation of the on-shell action. In section~\ref{sec:3}, we give analytic semiclassical 
AdS solutions with non-zero expectation values of the stress-energy tensor within perturbation. 
In section~\ref{sec:4}, we evaluate the free energy of the analytic solutions based on the prescription in section~\ref{sec:2}. 
Section~\ref{sec:5} is devoted to summary and discussions. The notation and conventions essentially follow our previous work~\cite{Ishibashi:2023luz}.

%%%%%%%%%%%%%%%%%
\section{The second variation of the on-shell action}\label{sec:2}
%%%%%%%%%%%%%%%%%

In this section, we evaluate the second-order variation of the effective action by using the AdS/CFT correspondence. We consider a $4$-dimensional AdS bulk spacetime with the metric 
\begin{align}
\label{metric}
ds_4^2 &=G_{MN}(X)dX^M dX^N \nonumber \\
&=\Omega^{-2}(z)dz^2+g_{\mu\nu}(z,x)dx^\mu dx^\nu \nonumber \\
&=\Omega^{-2}(z)(dz^2+\tilde{g}_{\mu\nu}(z,x)dx^\mu dx^\nu) \,, 
\end{align} 
where $X^M=(z, x^\mu)$ and $\Omega$ is a conformal factor which vanishes on the AdS conformal boundary at $z=0$.  
The conformal boundary metric $\bdyg_{\mu\nu}$ is defined by 
\begin{align}
\label{conformal_metric}
\bdyg_{\mu\nu}(x):=\lim_{z\to 0}\Omega^2(z)G_{\mu \nu}(z,\,x)=\lim_{z\to 0}\tilde{g}_{\mu\nu}(z,\,x) \,. 
\end{align} 

%%%%%%%%%%%%%%%%%%%%%%%%%%%%%%%%%%%%%%%
\vspace{-5mm} 
\begin{figure}[htbp]
  \begin{center}
\includegraphics[width=125mm]{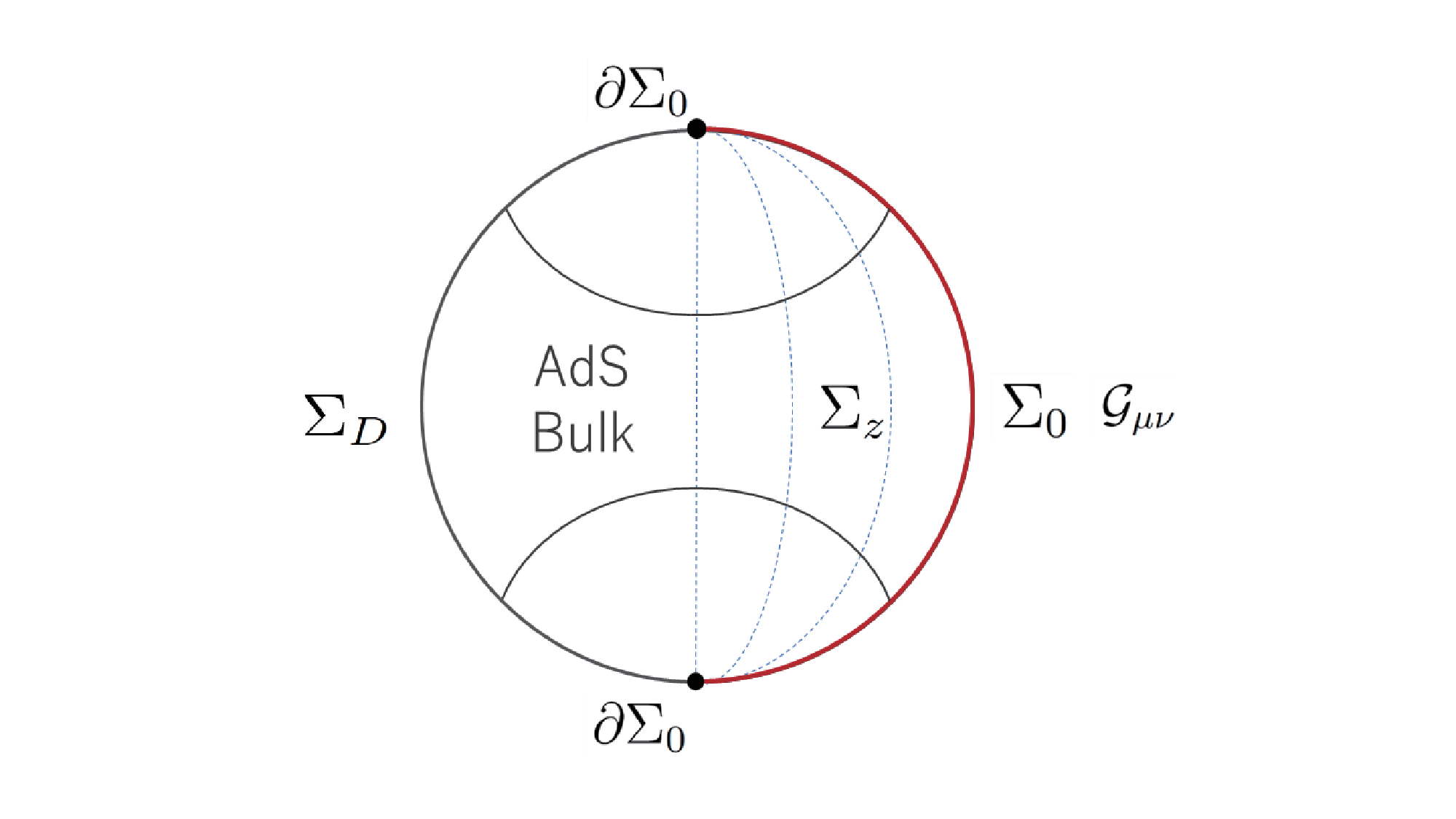}
  \end{center}
  \caption{{\small A time-slice of the (conformally compactified) AdS bulk spacetime foliated by $z=const$. hypersurfaces $\Sigma_z$ (denoted by the {\it dotted curve}), each of which itself is an asymptotically AdS spacetime one dimensional lower than the bulk AdS. The conformal boundary of the bulk AdS is divided into the left-part $\Sigma_D$ and the right-part $\Sigma_0$, and these two are matched at the corner $\partial \Sigma_0$. On the-right part $\Sigma_0$, the boundary metric is supposed to satisfy the holographic semiclassical Einstein equations, or in other words the mixed boundary condition is imposed on the bulk metric. For definiteness we assume that on the left-part $\Sigma_D$, the Dirichlet boundary conditions are imposed on the bulk metric. When $\Sigma_0$ includes a $3$-dimensional boundary black hole, the bulk spacetime also includes a $4$-dimensional black hole with a horizon $H$ inside the bulk. The two {\it hyperbolic curves} denote one of the possible bulk horizons. 
}}
  \label{conformal_ST}
 \end{figure} 
 %%%%%%%%%%%%%%%%%%%%%%%%%%%%%%%%%%%%%%

We assume that each $z= const.$ hypersurface---denoted by $\Sigma_z$---with the metric $\tilde{g}_{\mu \nu}(z,x)$ is asymptotically AdS$_3$ and that the AdS$_4$ bulk spacetime is foliated by a family of $\Sigma_z$, as shown in Fig.~\ref{conformal_ST}. Then, the limit hypersurface $\Sigma_0:=\lim_{z\to 0}\Sigma_z$ is a portion of the conformal boundary $\partial M$. 
%The boundary $\partial \Sigma_0$ of the $3$-dimensional spacetime $\Sigma_0$ corresponds to a $2$-dimensional AdS. 
We are concerned with the dynamics of $\Sigma_0$ which, in our setup, satisfies the holographic semiclassical Einstein equations. 
Following~\cite{Ishibashi:2023luz}, we shall impose the Dirichlet boundary condition at the other part of the conformal 
boundary $\Sigma_D:=\partial M \setminus \{ \Sigma_0 \cup \partial \Sigma_0 \}$~(see Fig.~\ref{conformal_ST}). 
The semiclassical Einstein equations are represented as a mixed boundary condition at $\Sigma_0$ for the 
bulk metric $G_{MN}$. Depending on the geometry of $\Sigma_0$, (e.g., when $\Sigma_0$ includes a $3$-dimensional black hole), the bulk spacetime may admit an inner boundary $H$ (e.g., the horizon of a $4$-dimensional bulk black hole or black string). See Fig.~\ref{conformal_ST}. 

The total effective action ${\mathcal S}$ for a $3$-dimensional semiclassical problem is constructed by 
the $3$-dimensional Einstein-Hilbert action ${\mathcal S}_\text{EH}$, the $2$-dimensional Gibbons-Hawking 
%term ${\mathcal S}_\text{GH}$, the $2$-dimensional counter term ${\mathcal S}_\text{ct}$, and the $4$-dimensional source 
%term $\Gamma$, which gives the vacuum stress-energy tensor $\Exp{\calT_{\mu\nu}}$ as   
term ${\mathcal S}_\text{GH}$, the $2$-dimensional counter term ${\mathcal S}_\text{ct}$,
and the effective action $\Gamma$ for $3$-dimensional CFT as
\begin{align}
\label{eff_action}
{\mathcal S}={\cal S}_\text{EH}+{\mathcal S}_\text{GH}+{\mathcal S}_\text{ct}+\Gamma \,, 
\end{align} 
where 
\begin{align}
\label{EH_action}
{\mathcal S}_\text{EH}=\frac{1}{16\pi G_3}
\int_{\Sigma_0} d^3x\sqrt{-\bdyg} \left(\bdyR-2\Lambda_3\right) \,,  
\end{align} 
with $G_3$, $\bdyR$, and $\Lambda_3$ being, respectively, the $3$-dimensional gravitational constant, 
the scalar curvature, and the cosmological constant on $(\Sigma_0, \bdyg_{\mu\nu})$, and where 
$\Gamma$ gives ries to the expectation value of stress-energy tensor for CFT: 
\begin{equation}
\Exp{\calT_{\mu\nu}} = - \dfrac{2}{ \sqrt{- \bdyg} } \,
\dfrac{ \delta \Gamma }{ \delta \bdyg^{\mu\nu} } \,. 
\label{def:SE:CFT3}
\end{equation}

According to the AdS/CFT correspondence, the effective action $\Gamma$ for CFT in (\ref{eff_action}) is given in terms of the bulk gravity dual. More precisely, $\Gamma$ is identified with the on-shell value of the bulk action $S_{\text{bulk}}$, composed of the $4$-dimensional Einstein-Hilbert action $S_\text{EH}$, the $3$-dimensional Gibbons-Hawking term $S_\text{GH}$, and the counter term $S_\text{ct}$, as 
\begin{align}
\label{bulk_action}
%& \Gamma=S_\text{EH}+S_\text{GH}+S_\text{ct}, \nonumber \\
& S_{\text{bulk}} =S_\text{EH}+S_\text{GH}+S_\text{ct}, \nonumber \\
& S_\text{EH}=\frac{1}{16\pi G_4}\int d^4X\sqrt{-G}\left(R(G)+\frac{6}{L^2}\right), \nonumber \\
& S_\text{GH}=\frac{1}{8\pi G_4}\int d^3x\sqrt{-g}\,K, \nonumber \\
& S_\text{ct}=-\frac{1}{16\pi G_4}\int d^3x\sqrt{-g}\left(\frac{4}{L}+LR(g) \right) \,, 
\end{align}
where $G_4$ and $L$ denote the $4$-dimensional gravitational constant and the curvature length, 
respectively, and where $R(G)$, $R(g)$ the scalar curvature of the bulk metric $G_{MN}$ and that of the induced metric $g_{\mu\nu}$ on $\Sigma_z$, respectively. 
Here, the extrinsic curvature $K_{\mu\nu}$ is defined by 
\begin{align}
\label{extrinsic_curvature}
K_{\mu\nu}=-\frac{\Omega}{2}\p_z g_{\mu\nu} \,. 
\end{align}
Since our effective action~(\ref{eff_action}) includes the Einstein-Hilbert term (\ref{EH_action}) on the conformal boundary~$\Sigma_0$, the conformal boundary metric $\bdyg_{\mu\nu}$ becomes dynamical~\cite{Compere:2008us}. Therefore, by varying the bulk metric $G_{MN}$ and $\bdyg_{\mu\nu}$ independently, we can obtain both the bulk Einstein equations and the boundary semiclassical Einstein equations: 
\begin{align}
\label{Basic_Eq}
& R_{MN}-\frac{1}{2}R\,G_{MN}-\frac{3}{L^2}G_{MN}=0, \nonumber \\
&{\bdyR}_{\mu\nu}-\frac{\bdyR}{2}\bdyg_{\mu\nu}-\frac{1}{\ell^2}\bdyg_{\mu\nu}-8\pi G_3
\pExp{\calT_{\mu\nu}}=0 \,, 
\end{align}
where $\ell$ is the curvature length $\ell^2=-1/ \Lambda_3$ and $\pExp{\calT_{\mu\nu}}$ is given by (\ref{def:SE:CFT3}). 

As shown in \cite{Ishibashi:2023luz}, the solution is obtained perturbatively by expanding 
the conformal (unphysical) metric $\tilde{g}_{\mu\nu}(z,\,x)$ as 
\begin{align}
\label{bulk_metric_series}
\tilde{g}_{\mu\nu}(z,\,x)={\Bar g}_{\mu\nu}(x)+\epsilon h_{\mu\nu}(z,\,x)+O(\epsilon^2) \,, 
\end{align}
%where $\epsilon$ is an infinitesimally small parameter and ${\Bar g}_{\mu\nu}(x)$ is the background 
where $\epsilon$ is an infinitesimally small parameter and
${\Bar g}_{\mu\nu}(x) = \Bar{\bdyg}_{\mu\nu}(x)$ is the background 
boundary metric: %BTZ black hole~(the global AdS$_3$) metric
\begin{align}
\label{BTZ_metric}
%ds_3^2
d\Bar{s}_3^2
=-\frac{f}{u}dt^2+\frac{\ell^2}{4u^2f}du^2+\frac{\ell^2d\varphi^2}{u}, \qquad f :=1-8\pi G_3{\cal M}\,u
\end{align}
with some constant $\cal M$, satisfying   
\begin{align}
\label{BTZ_Ricci}
%\bdyR_{\mu\nu}=-\frac{2}{\ell^2}\bdyg_{\mu\nu}.  
\Bar{\bdyR}_{\mu\nu}
=-\frac{2}{\ell^2} \Bar{\bdyg}_{\mu\nu} \,. 
\end{align} 
Here, the case ${\cal M}>0$ corresponds to the BTZ metric if $\varphi$ is $2\pi$-periodic, while $8\pi G_3{\cal M}=-1$ to the global AdS$_3$ metric. Note that the case ${\cal M}=0$ corresponds to (a locally) AdS$_3$ in the Poincare chart and case ${\cal M} <0$ (but $8\pi G_3{\cal M} \neq -1$) to (a locally) AdS$_3$ with a conical singularity (if $\varphi$ is $2\pi$-periodically identified), and in what follows, we do not consider these two cases.   

By using eqs.~(\ref{Basic_Eq}) and (\ref{BTZ_Ricci}), the conformal factor $\Omega$ is determined by 
\begin{align}
\label{conformal}
\Omega(z)=\frac{\ell}{L}\sin\frac{z}{\ell} \,, 
\end{align}
where $\Sigma_0$ and $\Sigma_D$ in Fig.~\ref{conformal_ST} are located at $z=0$ and $z=\pi\ell$, respectively. 
Note that in the unperturbed case, $\epsilon=0$, the expectation value of the stress-energy tensor 
$\pExp{\calT_{\mu\nu}}$ vanishes, and therefore the semiclassical equations in eqs.~(\ref{Basic_Eq}) are trivially satisfied.

Now we holographically evaluate the second order variation of the effective action $\Gamma$ in~(\ref{eff_action}) by inspecting the action for the gravity dual~(\ref{bulk_action}). Let us first examine the bulk Einstein-Hilbert action, 
% In terms of the conformal metric $\tilde{G}_{MN}\equiv \Omega^2\,G_{MN}$, $S_\text{EH}$ is rewritten by 
\begin{align} \label{SEH4}
%S_\text{EH}=\frac{1}{16\pi G_4}\int d^4X\left[\sqrt{-\tilde{g}}\left(\frac{\tilde{R}(\tilde{G})}{\Omega^2}
S_\text{EH}=\frac{1}{16\pi G_4}\int d^4X\left[\sqrt{-\tilde{g}}\left(
\frac{{R}(\tilde{G})}{\Omega^2}
-\frac{12\Omega'^2}{\Omega^4}
+\frac{6\Omega''}{\Omega^3}+\frac{6}{L^2\Omega^4}\right)+\frac{6\Omega'}{\Omega^3}(\sqrt{-\tilde{g}})'    \right] \,,  
\end{align}
where ${R}(\tilde{G})$ is the scalar curvature of the conformal metric $\tilde{G}_{MN} := \Omega^2\,G_{MN}$ and the {\it prime} denotes 
the derivative with respect to $z$. 

As our variation, we consider the tensor-type perturbations of the bulk metric which satisfy $h_{zz}=h_{z\nu}=0$, and  
\begin{align}
\label{TT_condition}
{h_\nu}^\nu=h_{\mu\nu}\Bar{g}^{\mu\nu}=0, \qquad \Bar{D}^\nu h_{\nu\mu}=0 \,,  
\end{align}
where $\Bar{D}_\mu$ is the covariant derivative with respect to the unperturbed boundary metric~$\Bar{g}_{\mu\nu}$.   

% 
%In terms of the conformal metric $\tilde{G}_{MN}\equiv \Omega^2\,G_{MN}$, $S_\text{EH}$ is rewritten by 
%\begin{align}
%%S_\text{EH}=\frac{1}{16\pi G_4}\int d^4X\left[\sqrt{-\tilde{g}}\left(\frac{\tilde{R}(\tilde{G})}{\Omega^2}
%S_\text{EH}=\frac{1}{16\pi G_4}\int d^4X\left[\sqrt{-\tilde{g}}\left(
%\frac{{R}(\tilde{G})}{\Omega^2}
%-\frac{12\Omega'^2}{\Omega^4}
%+\frac{6\Omega''}{\Omega^3}+\frac{6}{L^2\Omega^4}\right)+\frac{6\Omega'}{\Omega^3}(\sqrt{-\tilde{g}})'    \right] \,,  
%\end{align}
%where ${R}(\tilde{G})$ is the scalar curvature of the metric $\tilde{G}_{MN}$ and the {\it prime} denotes 
%the derivative with respect to $z$.

% \textcolor{magenta}{以下, 摂動の次数を表すため, 次の記法を用いる:
%
% \begin{align}
%  & Q(\epsilon)
%  = \Bar{Q} + \delta Q + \frac{1}{2}\, \delta^{2} Q + \cdots
% & & \left( \delta^{n} Q
% := \epsilon^n\, \frac{d^nQ}{d\epsilon^n}\, \bigg|_{\epsilon=0}
%  \right)
%  ~,
%\end{align}
%
% }　記法注
By using eqs.~(\ref{TT_condition}) and (\ref{first_vari_R}), 
it is easily checked that the first variation of the bulk action (\ref{SEH4}) vanishes. With the help of the formulas~(\ref{second_variations}) and 
(\ref{second_Ricci_scalar}), we obtain the second variation of the bulk action as 
\begin{align}
\label{second_variation_bulk}
%& \delta^{(2)}S_\text{EH}=\frac{\epsilon^2}{16\pi G_4}\int d^4X\sqrt{-\Bar{g}}
\delta^{2}S_\text{EH} 
 &=\frac{\epsilon^2}{16\pi G_4}\int d^4X\sqrt{-\Bar{g}}
\Biggl[\frac{3}{L^2\Omega^4}h_{\mu\nu}h^{\mu\nu}-\frac{3\Omega'}{\Omega^3}(h_{\mu\nu}h^{\mu\nu})'
\nonumber \\
&{} \quad +\frac{1}{\Omega^2}\Biggl\{-\frac{h_{\mu\nu}h^{\mu\nu}}{\ell^2}
+\frac{1}{2}h^{\mu\nu}(\Bar{D}^2h_{\mu\nu}+h_{\mu\nu}'')+\frac{3}{4}(h_{\mu\nu}h^{\mu\nu})''
+\Bar{D}_\mu V^\mu)
\Biggr\}\Biggr] \,, 
\end{align}
where $V^\mu$ is defined by 
\begin{align}
\label{def_V}
V^\mu:=\frac{3}{4}\Bar{D}^\mu(h^{\alpha\beta}h_{\alpha\beta})-\Bar{D}_\nu(h^{\mu\alpha}{h_\alpha}^\nu) \,. 
\end{align}
Using the following relations
\begin{align}
\label{Omega_relation}
\Omega''=-\frac{\Omega}{\ell^2} \,, \quad 
\frac{1-L^2\Omega'^2}{L^2\Omega^2} =\frac{1}{\ell^2} \,, \quad 
h'_{\mu\nu}h^{\mu\nu}=\frac{1}{2}(h_{\mu\nu}h^{\mu\nu})' \,, 
\end{align}
we can rewrite (\ref{second_variation_bulk}) as  
\begin{align}
\label{second_variation_bulk2}
%&\delta^{(2)}S_\text{EH}
 \delta^{2} S_\text{EH}
&=\frac{\epsilon^2}{32\pi G_4}\int d^4X\frac{\sqrt{-\Bar{g}}}{\Omega^2}h^{\mu\nu}
\Biggl\{\Omega^2\left(\frac{h_{\mu\nu}'}{\Omega^2} \right)'+\left({\Bar D}^2+\frac{2}{\ell^2}\right)h_{\mu\nu}\Biggr\}
\nonumber \\
&{} +\frac{\epsilon^2}{16\pi G_4}\int d^4X\sqrt{-\Bar{g}}
\biggl[\left\{\frac{3}{4\Omega^2}(h^{\mu\nu}h_{\mu\nu})'-\frac{\Omega'}{\Omega^3}h^{\mu\nu}h_{\mu\nu}\right\}'
+\frac{1}{\Omega^2}\Bar{D}_\mu V^\mu\Biggr] \nonumber \\
&=\frac{\epsilon^2}{16\pi G_4}\int d^4X\sqrt{-\Bar{g}}
\biggl[\left\{\frac{3}{4\Omega^2}(h^{\mu\nu}h_{\mu\nu})'-\frac{\Omega'}{\Omega^3}h^{\mu\nu}h_{\mu\nu}\right\}'
+\frac{1}{\Omega^2}\Bar{D}_\mu V^\mu\Biggr] \,, 
\end{align}
where in the second equality, we have used the perturbed bulk equation derived in Ref.~\cite{Ishibashi:2023luz},
\begin{align}
\label{EOM_per}
h_{\mu\nu}''-\frac{2\Omega'}{\Omega}h_{\mu\nu}'+\left({\Bar D}^2+\frac{2}{\ell^2}\right)h_{\mu\nu}=0 \,. 
\end{align}
As expected, only the surface terms are left on the evaluation of $\delta^{2}S_\text{EH}$ under the on-shell condition.  
 
Similarly, we obtain the second variations of $S_\text{GH}$ and $S_\text{ct}$  in~(\ref{bulk_action}) with respect to the tensor-type perturbations~(\ref{TT_condition}) as 
\begin{align} 
\label{second_variation_GH}
%\delta^{(2)} S_\text{GH}=\frac{\epsilon^2}{16\pi G_4}\int d^3x\frac{\sqrt{-\Bar{g}}}{\Omega^2}
\delta^{2} S_\text{GH}=\frac{\epsilon^2}{16\pi G_4}\int_{\Sigma_0}\! d^3x\frac{\sqrt{-\Bar{g}}}{\Omega^2}
\left[(h^{\mu\nu}h_{\mu\nu})'-\frac{3\Omega'}{\Omega}h^{\mu\nu}h_{\mu\nu}\right] \,,  
\end{align}
\begin{align} 
\label{second_variation_ct}
%\delta^{(2)} S_\text{ct}=
{\delta^{2}} S_\text{ct}=
-\frac{\epsilon^2L}{16\pi G_4 }\int_{\Sigma_0} \! d^3x\frac{\sqrt{-\Bar{g}}}{\Omega}\left[\frac{1}{2}h^{\mu\nu}
\left(\Bar{D}^2+\frac{2}{\ell^2}\right)h_{\mu\nu}
-\left(\frac{1}{\ell^2}+\frac{2\Omega'^2}{\Omega^2}\right)h^{\mu\nu}h_{\mu\nu}
+\Bar{D}_\mu V^\mu\right] \,, 
\end{align}
where we have used $R(g)=\Omega^2\bdyR$ and the second variation of 
$\bdyR$ in (\ref{second_bdyR}).  

Combining eqs.~(\ref{second_variation_bulk}), (\ref{second_variation_GH}), and (\ref{second_variation_ct}), 
%we obtain the second variation of the action $\Gamma$ in (\ref{bulk_action})  
we obtain the second variation of the {effective}
action $\Gamma$ in (\ref{bulk_action})  
\begin{align} 
\label{second_variation_Gamma}
%& \delta^{(2)} \Gamma=-\frac{\epsilon^2L}{16\pi G_4}\int d^3x\frac{\sqrt{-\Bar{g}}}{\Omega}
\delta^{2} \Gamma 
& =-\frac{\epsilon^2L}{16\pi G_4}\int_{\Sigma_0} \!d^3x\frac{\sqrt{-\Bar{g}}}{\Omega}
\Biggl[\frac{1}{2}h^{\mu\nu}\left(\Bar{D}^2+\frac{2}{\ell^2}  \right)h_{\mu\nu}
-\frac{1}{4L\Omega}(h^{\mu\nu}h_{\mu\nu})' \nonumber \\
&{}\quad +\Biggl\{\frac{2\Omega'}{L\Omega^2}
-\left(\frac{1}{\ell^2}+\frac{2\Omega'^2}{\Omega^2} \right)\Biggr\}h^{\mu\nu}h_{\mu\nu} \Biggr]
%+\epsilon^2({\cal I}_l+{\cal I}_\text{B}+{\cal I}_\text{c}) \nonumber \\
+\epsilon^2({\mathcal{I}_D } +{\cal I}_\text{B}+{\cal I}_\text{c}) \nonumber \\
&=-\frac{\epsilon^2L}{16\pi G_4}\int_{\Sigma_0} \! d^3x\frac{\sqrt{-\Bar{g}}}{\Omega}
\Biggl[-\frac{1}{2}h^{\mu\nu}\left(\frac{\p}{\p z}-\frac{1}{L\Omega}\right)h'_{\mu\nu}
-\frac{L\Omega h^{\mu\nu}h'_{\mu\nu}}{\ell^2(1+L\Omega')}
-\frac{L^2\Omega^2 h^{\mu\nu}h_{\mu\nu}}{\ell^4(1+L\Omega')^2}\Biggr] \nonumber \\
&{} \quad +\epsilon^2({\cal I}_\text{B}+{\cal I}_\text{c}) \,, 
\end{align}
where 
\begin{align}
\label{surface_term1}
%{\cal I}_l=\frac{1}{16\pi G_4}\int_{\Sigma_D}\frac{\sqrt{-\Bar{g}}}{\Omega^2}
\mathcal{I}_D&=\frac{1}{16\pi G_4}\int_{\Sigma_D}\frac{\sqrt{-\Bar{g}}}{\Omega^2}
\left\{\frac{3}{4}(h^{\mu\nu}h_{\mu\nu})'-\frac{\Omega'}{\Omega}h^{\mu\nu}h_{\mu\nu}  \right\} \,,   
\end{align}
\begin{align}
\label{surface_termB}
 {\cal I}_\text{B} &=\frac{1}{16\pi G_4}\int d^4X\frac{\sqrt{-\Bar{g}}}{\Omega^2}\Bar{D}_\mu V^\mu
\nonumber \\ 
% &=\frac{1}{16\pi G_4}\int_{\partial M/\Sigma_D\cup \Sigma_0} \frac{\sqrt{-\Bar{h}}}{\Omega^2}\Bar{n}_\mu V^\mu
&=\frac{1}{16\pi G_4}\int_{\partial \Sigma_0} \frac{\sqrt{-\Bar{h}}}{\Omega^2}\Bar{n}_\mu V^\mu
+\frac{1}{16\pi G_4}\int_H \frac{\sqrt{-\Bar{h}}}{\Omega^2}\Bar{n}_\mu V^\mu \,, 
\end{align}
\begin{align}
\label{surface_termC}
{\cal I}_\text{c}
&=-\frac{L}{16\pi G_4}\int_{\Sigma_0}\frac{\sqrt{-\Bar{g}}}{\Omega}\Bar{D}_\mu V^\mu
 =-\frac{L}{16\pi G_4}\int_{\partial \Sigma_0}\frac{\Bar{n}_\mu V^\mu}{\Omega} \,, 
\end{align}
where $\Bar{n}^\mu$ is the unit normal vector to the boundary surfaces, $\partial M \setminus \{\Sigma_D\cup \Sigma_0 \} $, $\partial \Sigma_0$, and $H$. Here, $H$ denotes an inner boundary, such as the horizon of a black hole, if exists. Note that the boundary integral of ${\cal I}_{\rm B}$ is 
performed first inside the bulk and then is taken the limit toward $\partial \Sigma_0$, whereas the integral of ${\cal I}_{\rm c}$ should be taken 
on the boundary $\Sigma_0$ and taken the limit to $\partial \Sigma_0$.

%In the second equality of Eq.~(\ref{second_variation_Gamma}), we used the fact ${\cal I}_l=0$ by the Dirichlet 
In the second equality of eq.~(\ref{second_variation_Gamma}),
we used the fact $\mathcal{I}_D =0$ by the Dirichlet 
boundary condition imposed on $\Sigma_D$, and the 
derivative operator $\Bar{D}^2$ is eliminated by eq.~(\ref{EOM_per}), and used the second 
equation in (\ref{Omega_relation}).

Near the conformal boundary $\Sigma_0$, $h_{\mu\nu}$ can be expanded as a series in $z$ as 
\begin{align}
\label{series_h}
h_{\mu\nu}(z,\,x)=h^{(0)}_{\mu\nu}(x)+z^2h^{(2)}_{\mu\nu}(x)+z^3h^{(3)}_{\mu\nu}(x)+\cdots \,. 
\end{align}
Substituting (\ref{series_h}) into the square brackets in the second equality of eq.~(\ref{second_variation_Gamma}), we obtain 
\begin{align}
\label{bracket}
%& \delta^{(2)} \Gamma=\frac{3\epsilon^2L^2}{32\pi G_4}\int d^3x\sqrt{-\bm{g}}
& \delta^{2} \Gamma
=\frac{3\epsilon^2L^2}{32\pi G_4}\int_{\Sigma_0} \!d^3x\sqrt{ - {\Bar{g}} }
\Biggl[h_{(0)}^{\mu\nu}(x)h^{(3)}_{\mu\nu}(x)+O(z)  \Biggr]+{\cal I}_B+{\cal I}_c \,, % \nonumber \\
%&=\frac{\epsilon}{2}\int d^3x\sqrt{-\bm{g}}\,h_{(0)}^{\mu\nu}(x)\delta\Exp{\calT_{\mu\nu}}+{\cal I}_B+{\cal I}_c, 
% &=\frac{ {- \epsilon} }{ \cancel{2}} \int d^3x\sqrt{ - {\Bar{g}} }\,h_{(0)}^{\mu\nu}(x)\delta\Exp{\calT_{\mu\nu}}+{\cal I}_B+{\cal I}_c, 
\end{align}
where, noting the fact that the on-shell $h^{(3)}_{\mu\nu}$ contains the first order terms of $h^{(0)}_{\mu\nu}$, we have used the formula: 
\begin{align}
\delta\Exp{\calT_{\mu\nu}}=\frac{3\epsilon L^2}{16\pi G_4}h^{(3)}_{\mu\nu}(x) \,. 
\end{align}
In the limit $z\to 0$, the surface term ${\cal I}_c$ diverges, and we should discard this term when evaluating the free energy 
in the next section. ${\cal I}_B$ is also a surface term perpendicular to each $z=const.$ surface, or 
on the horizon $H$. In the spirit of the AdS/CFT correspondence, we should also discard this term because 
$\Gamma$ should be a functional of the AdS boundary $\Sigma_0$. 
% So, hereafter, we discard both of the surface terms.   

%%%%%%%%%%%%%%%%%%%%%%%%%%%
\section{The linear solutions}\label{sec:3}
%%%%%%%%%%%%%%%%%%%%%%%%%%%
In this section, we construct two regular static solutions satisfying both the bulk Einstein equations and 
the boundary semiclassical Einstein equations~(\ref{Basic_Eq}). We make the following ansatz for separation of variables for the 
perturbed metric $h_{\mu\nu}(z,\,x)$ in eq.~(\ref{bulk_metric_series}) as 
\begin{align}
\label{separation}
h_{\mu\nu}(z,\,x)=\xi(z)H_{\mu\nu}(x) \,. 
\end{align}
Then, the perturbed bulk equations~(\ref{EOM_per}) are decomposed into the $3$-dimensional part  
\begin{align}
\label{3-dim_part}
\Bar{D}^2H_{\mu\nu}+\frac{2}{\ell^2}H_{\mu\nu}=m^2H_{\mu\nu} \,, 
\end{align}
and the radial part~\cite{Ishibashi:2023luz}
\begin{align}
\label{radial_part}
\left(\frac{d^2}{dz^2}-2\frac{\Omega'}{\Omega}\frac{d}{dz}+m^2\right)\xi(z)=0
\end{align}
with a separation constant $m^2$. 

We express our perturbation variable $H_{\mu\nu}$ in eq.~(\ref{separation}) in terms of three functions $(T, Y, U)$ of $u$ as follows,  
% around the background metric~(\ref{BTZ_metric}) is 
\begin{align}
\label{per_BDmetric}
%& ds^2=(\bdyg_{\mu\nu}+\epsilon H_{\mu\nu})dx^\mu dx^\nu \nonumber \\
{ds_3}^2 
&=({\Bar{\bdyg}}_{\mu\nu}+\epsilon H_{\mu\nu})dx^\mu dx^\nu \nonumber \\
&=-\frac{f}{u}(1+\epsilon T(u))dt^2+\frac{1}{u}(1+\epsilon Y(u))dy^2
+\frac{\ell^2}{4u^2f}(1+\epsilon U(u))du^2 \,, 
\end{align} 
where $f(u)=1-8\pi G_3{\cal M}\,u$ as defined before, $y$ is related to the angular coordinate $\varphi$ as $y=\ell \varphi$, and 
the $2$-dimensional boundary~($u=0$) corresponds to $\partial\Sigma_0$ in Fig.~\ref{conformal_ST}. 
The transvers-traceless condition~(\ref{TT_condition}) reduces to 
\begin{align}
\label{TT_condition2}
& U+T+Y=0 \,, \nonumber \\
& \left(u\frac{d}{du}-\frac{3}{2}\right)U+\frac{uf'}{2f}(U-T)=0 \,.
\end{align}
Combining eq. (\ref{3-dim_part}) with eqs.~(\ref{TT_condition2}), we obtain the following master equation 
\begin{align}
\label{master_U}
\left(\frac{d^2}{du^2}-\frac{2}{uf}\frac{d}{du}-\frac{\hat{m}^2-8}{4u^2f}\right)U=0 \,,  
\end{align}  
where $\hat{m}^2=\ell^2m^2$. 

The general solutions to (\ref{master_U}) can be obtained in terms of the hypergeometric functions $F(\alpha,\beta, \gamma;x)$. Since the expression of the solutions depends on the chart chosen, we denote with superscripts ${}^{({\rm global})}$ and ${}^{({\rm BTZ})}$ the solutions and related quantities in the global AdS$_3$ chart and in the BTZ chart, respectively. The general solutions are given by    
\begin{align}
\label{sol_global}
U^{({\rm global})}(u) 
&=C_1^{({\rm global})}u^{\frac{3-p}{2}}
F\left(\frac{1-p}{2}, \,\frac{3-p}{2},\,1-p;\,-u \right)
\nonumber \\
&+C_2^{({\rm global})}u^{\frac{3+p}{2}}
F\left(\frac{1+p}{2} \,, \,\frac{3+p}{2},\,1+p;\,-u \right) \,,  
\end{align}
\begin{align}
\label{sol_BTZ}
U^{(\text{BTZ})}(u)
& =\frac{C_1^{(\text{BTZ})}}{u_H}u^{\frac{3-p}{2}}
F\left(\frac{1-p}{2} \,, \,\frac{3-p}{2},\,1-p;\,-\frac{u}{u_H} \right)
\nonumber \\
&+\frac{C_2^{(\text{BTZ})}}{u_H}u^{\frac{3+p}{2}}
F\left(\frac{1+p}{2}, \,\frac{3+p}{2},\,1+p;\,-\frac{u}{u_H} \right) \,, 
\end{align}
%where $u_H:=\frac{1}{8G_3{\cal M}}$ and $p=\sqrt{1+\hat{m}^2}$. 
where $u_H:= {1}/{8G_3{\cal M}}$ and $p {\,:=}\sqrt{1+\hat{m}^2}$. 
%, and $U^{({\rm g})}(u)$ and $U^{(\text{BTZ})}(u)$ represent the solutions of eq.~(\ref{master_U}) in the background of the global AdS$_3$ and the BTZ black hole, respectively. 
By imposing the regularity at the center, $u=\infty$, for the global AdS$_3$ case, and at the horizon, $u=u_H$, for the BTZ case, we obtain the following relation between the coefficients 
$C_1^{({\rm global})}$, $C_2^{({\rm global})}$, and $C_1^{(\text{BTZ})}$, $C_2^{(\text{BTZ})}$ as 
\begin{align}
\label{Coeff_relation_g}
\frac{C_2^{({\rm global})}}{C_1^{({\rm global})}}=\frac{-1}{4^{p}}
\frac{\Gamma\left(1-\frac{p}{2} \right)\Gamma\left(3+\frac{p}{2} \right)}
{\Gamma\left(1+\frac{p}{2} \right)\Gamma\left(3-\frac{p}{2} \right)} \,, 
\end{align}
\begin{align}
\label{Coeff_relation_BTZ}
\frac{C_2^{(\text{BTZ})}}{C_1^{(\text{BTZ})}}=\frac{-1}{(4u_H)^p}
\frac{\Gamma\left(1-\frac{p}{2} \right)\Gamma\left(3+\frac{p}{2} \right)}
{\Gamma\left(1+\frac{p}{2} \right)\Gamma\left(3-\frac{p}{2} \right)} \,. 
\end{align}

%%%%%%%%%%%%%%%%%%%%%%%%%%%%%%%%%%%%%%%
\begin{figure}[htbp]
  \begin{center}
\includegraphics[width=90mm]{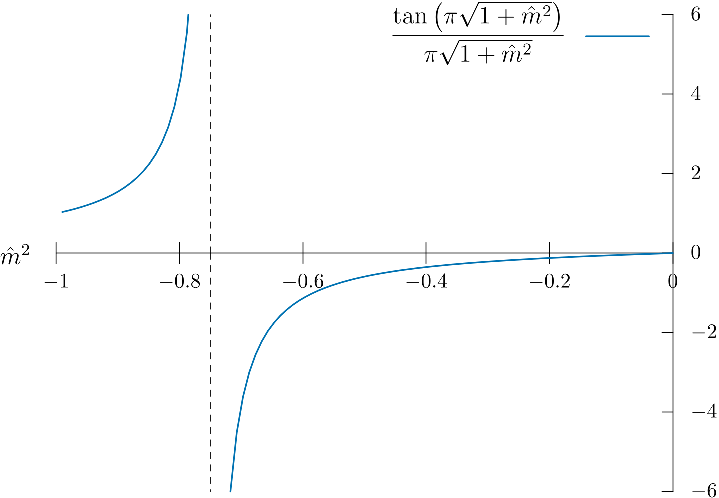}
  \end{center}
  \caption{{\small The plot of $\tan\pi\sqrt{1+\hat{m}^2}/\pi\sqrt{1+\hat{m}^2}$. When $\gamma_3>1$, there is only 
  one solution in the range $-1<\hat{m}^2<-3/4$. 
% When the gravitational constant on the boundary is zero, i.e., $G_3=0$~($\gamma_3=0$), $\hat{m}^2=0$ massless solutions also exist, although it is nothing but the background solution.
}}
  \label{fig_tan}
 \end{figure} 
 %%%%%%%%%%%%%%%%%%%%%%%%%%%%%%%%%%%%%%
As shown in Ref.~\cite{Ishibashi:2023luz}, the semiclassical solutions are determined by 
\begin{align}
\label{dimensionless_relation}
\gamma_3=\frac{\tan\pi\sqrt{1+\hat{m}^2}}{\pi\sqrt{1+\hat{m}^2}} = \dfrac{\tan \pi p}{\pi p} \,, 
\end{align} 
where $\gamma_3$ is the dimensionless parameter  
\begin{align}
\label{dimensionless_para}
\gamma_3:=\frac{G_3}{G_4}\frac{L^2}{\pi \ell} \,.
\end{align}
Under the Dirichlet boundary condition on $\Sigma_D$ and the mixed boundary condition on $\Sigma_0$, 
the non-trivial solution exists only when $\gamma_3>1$ in the range $-1<\hat{m}^2<-3/4$ ($0<p<1/2$). See Fig.~\ref{fig_tan}. 
Then, the ratio between $C_1$ and $C_2$ becomes negative, i.e.,  
\begin{align}
\label{sign_C1C2}
C_1C_2<0
\end{align}
in both the global AdS$_3$ case and the BTZ case.

%%%%%%%%%%%%%%%%%%%%%%%%%%%
\section{The boundary free energy}\label{sec:4}
%%%%%%%%%%%%%%%%%%%%%%%%%%%
To evaluate the total effective action~(\ref{eff_action}), one needs to derive the second variation 
of the boundary action ${\cal S}_\text{bdy}={\cal S}_\text{EH}+{\cal S}_\text{GH}+{\cal S}_\text{ct}$, combined with 
the bulk calculation~(\ref{bracket}).  
The $2$-dimensional GH term ${\cal S}_\text{GH}$ and the $2$-dimensional counter term ${\cal S}_\text{ct}$ 
are defined by 
\begin{align}
\label{two_boundary_terms}
%& {\cal S}_\text{GH}=\frac{1}{8\pi G_3}\int dtdy\sqrt{-\sigma}\,g^{ab}{\mathcal K}_{ab}, \quad a,\,b=t,\,y, 
 {\cal S}_\text{GH} &=\frac{1}{8\pi G_3}
 \int dt dy\sqrt{-\sigma}\,
 {\sigma}^{ab}{\mathcal K}_{ab} \Big|_{u=0} \,, \quad 
 \left( a,\,b=t,\,y \right) 
\nonumber \\
%& {\cal S}_\text{ct}=\frac{\alpha}{16\pi G_3}\int dtdy\sqrt{-\sigma}, 
 {\cal S}_\text{ct} &= \frac{\alpha}{16\pi G_3}
\int dtdy\sqrt{-\sigma} \Big|_{u=0} \,, 
\end{align} 
%where $\sigma=\det (g_{ab})$ and ${\mathcal K}_{ab}$ is the $2$-dimensional extrinsic curvature given by 
where $\sigma_{ab} := \bdyg_{ab}$ and ${\mathcal K}_{ab}$ is the $2$-dimensional extrinsic curvature given by 
\begin{align}
%{\mathcal K}_{ab}:=-\frac{1}{2N}\p_u g_{ab} = -\frac{u\sqrt{f}}{\ell \sqrt{1 +\epsilon U}}\p_u g_{ab}, 
{\mathcal K}_{ab}:=-\frac{1}{2N}\p_u {\sigma}_{ab}
= -\frac{u\sqrt{f}}{\ell \sqrt{1 +\epsilon U}}\p_u {\bdyg}_{ab} \,, 
\end{align}
%and $N$, $\alpha$ are the lapse function of the metric~(\ref{per_BDmetric}). Note that the parameter $\alpha$ is later chosen to eliminate the divergent terms in the total action~(\ref{eff_action}).  See, below (\ref{para_alpha}). 
and $N$ is the lapse function of the metric~(\ref{per_BDmetric}). Note that the parameter $\alpha$ is to be chosen so that the divergent terms in the total action~(\ref{eff_action}) are eliminated. See below (\ref{para_alpha}).  

The second variation of the boundary Einstein-Hilbert action ${\cal S}_\text{EH}$ can be also evaluated via 
the formulas (\ref{second_variations}) and (\ref{second_bdyR}) as 
\begin{align}
\label{second_variation_bdy}
%& \delta^{(2)}{\cal S}_\text{EH}=\frac{\epsilon^2}{16\pi G_3}\int_{\Sigma_0} d^3x\sqrt{-\Bar{g}}\Biggl[\frac{1}{2}h^{(0)\mu\nu}
{\delta^{2}} {\cal S}_\text{EH} &=\frac{\epsilon^2}{16\pi G_3}\int_{\Sigma_0} d^3x\sqrt{-\Bar{g}}\Biggl[\frac{1}{2}h^{(0)\mu\nu}
\left(\Bar{D}^2+\frac{2}{\ell^2}\right)h^{(0)}_{\mu\nu}+\Bar{D}_\mu V^\mu\Biggr] \nonumber \\
&=\frac{\epsilon^2}{32\pi G_3}\int_{\Sigma_0} d^3x\sqrt{-\Bar{g}}\,h^{(0)\mu\nu}
%\left(\Bar{D}^2+\frac{2}{\ell^2}\right)h^{(0)}_{\mu\nu}+\delta^{(2)}{\cal S}_\text{V}, 
\left(\Bar{D}^2+\frac{2}{\ell^2}\right)h^{(0)}_{\mu\nu} + {\delta^{2}} {\cal S}_\text{V} \,, 
\end{align} 
%where $\delta^{(2)}{\cal S}_\text{V}$ is defined on the $2$-dimensional timelike boundary $\partial \Sigma_0$ at $u=0$ as 
where $\Bar{\bdyg}_{\mu\nu}(x) = \Bar{g}_{\mu\nu}(x)$, $\delta \bdyg_{\mu\nu}(x) = h^{(0)}_{\mu\nu}(x) := \lim_{z \to 0} h_{\mu\nu}(z, x)$. 
Here note that ${\delta^{2}} {\cal S}_\text{V}$ is defined on the $2$-dimensional timelike boundary $\partial \Sigma_0$ at $u=0$ as 
\begin{align}
%\delta^{(2)}{\cal S}_\text{V}=\frac{\epsilon^2}{16\pi G_3}\int_{\partial \Sigma_0} 
\delta^{2}{\cal S}_\text{V}=\frac{\epsilon^2}{16\pi G_3}\int_{\partial \Sigma_0} 
dtdy\sqrt{-\Bar{\sigma}}\,\Bar{n}_\mu V^\mu \,, 
\end{align}
where $n^\mu$ is the unit outward vector normal to the boundary $\partial \Sigma_0$. 
Note also that 
% \textcolor{magenta}{ \small ($\partial \Sigma_0$ が $\Sigma_0$ 上の horizon を含まないなら, 
% (最終的に寄与しないことが分かるが)
in general there could be some contribution from ${H \cap \Sigma_0}$ to $\delta^2 {\cal S}_V$, but in the present case, we do not have such contributions.

From eqs.~(\ref{bracket}) and (\ref{second_variation_bdy}), we obtain the second variation of the total effective action~(\ref{eff_action}), 
\begin{align}
\label{second_variation_eff}
%& \delta^{(2)}{\mathcal S}=-\frac{\epsilon}{16\pi G_3}\int_{\Sigma_0} d^3x
 {\delta^{2}}{\mathcal S} &=-\frac{\epsilon}{16\pi G_3}\int_{\Sigma_0} d^3x
\sqrt{-\Bar{g}}h^{(0)\mu\nu}\Biggl[-\frac{\epsilon}{2}
\left(\Bar{D}^2+\frac{2}{\ell^2}\right)h^{(0)}_{\mu\nu}-8\pi G_3\delta\Exp{\calT_{\mu\nu}} \Biggr] \nonumber \\
%&+\delta^{(2)}{\cal S}_\text{V}
%+ \delta^{(2)}{\cal S}_\text{GH}+ \delta^{(2)}{\cal S}_\text{ct} \nonumber \\
&\quad +{\delta^{2}}{\cal S}_\text{V}
+ {\delta^{2}}{\cal S}_\text{GH} + {\delta^{2}}{\cal S}_\text{ct} \,. % \nonumber \\
%&
%\textcolor{magenta}{\xout{
%=\delta^{(2)}{\cal S}_\text{V}
%+ \delta^{(2)}{\cal S}_\text{GH}+ \delta^{(2)}{\cal S}_\text{ct}
%}}
% .
\end{align} 
The first and second order variations of ${\cal S}_\text{GH}$ and ${\cal S}_\text{ct}$ are obtained as 
\begin{align}
\label{ex_SGH_1}
%{\cal S}^{(1)}_\text{GH}=\frac{1}{8\pi G_3\ell}\int_{\partial \Sigma_0} 
{\delta \mathcal{S} }_\text{GH}
=\frac{ {\epsilon} }{8\pi G_3\ell}\int_{\partial \Sigma_0} 
dtdy\left[-f'U+\frac{f}{u}(uU'-2U) \right] \,,  
\end{align} 
\begin{align}
\label{ex_SGH_2}
%& {\cal S}^{(2)}_\text{GH}=\frac{1}{64\pi G_3\ell}\int_{\partial \Sigma_0}
{\delta^2 \mathcal{S} }_\text{GH}
&=\frac{{\, \epsilon^2} }{32\pi G_3\ell}\int_{\partial \Sigma_0}
 dtdy\Biggl[-(1+f)\{T^2+2T(U-Y)-(U-Y)(3U+Y)\}
\nonumber \\
&{} \qquad \qquad \qquad +4uf\{(T+U-Y)T'+(Y+U-T)Y'\}\Biggr] \,, 
\end{align}
and % \textcolor{magenta}{および,}
\begin{align}
\label{ex_Sct_1}
% {\cal S}^{(1)}_\text{ct}=\frac{\alpha}{16\pi \ell G_3}
 {\delta \mathcal{S} }_\text{ct}
 =\frac{{\epsilon}\, \alpha}{16\pi \ell G_3}
 \int_{\partial \Sigma_0} dtdy\frac{\sqrt{f}}{u}U \,, 
\end{align}
\begin{align}
\label{ex_Sct_2}
% {\cal S}^{(2)}_\text{ct}=-\frac{\alpha}{64\pi \ell G_3}
 {\delta^2 \mathcal{S} }_\text{ct}
 =-\frac{ {\, \epsilon^2}\, \alpha}{32\pi \ell G_3}
 \int_{\partial \Sigma_0} dtdy\frac{\sqrt{f}}{u}(T-Y)^2 \,, 
\end{align}
where %$\partial \Sigma_0$ is the $u=0$ AdS boundary and 
we have used  (\ref{TT_condition2}) in eqs.~(\ref{ex_SGH_1}), (\ref{ex_Sct_1}), and $f'=(f-1)/u$ in eq.~(\ref{ex_SGH_2}). Since $U$ behaves near $u=0$ as 
$U\sim u^{(3-\sqrt{1+\hat{m}^2})/2}$ as seen from eqs.~(\ref{sol_global}) and (\ref{sol_BTZ}), and 
$-1<\hat{m}^2<-3/4$, we find 
\begin{align}
%{\cal S}^{(1)}_\text{GH}= {\cal S}^{(1)}_\text{ct}=0. 
 { \delta \mathcal{S} }_\text{GH}
= { \delta \mathcal{S} }_\text{ct}=0 \,. 
\end{align}
Therefore the surface terms at $O(\epsilon)$ do not appear when one evaluates the on-shell action.  

At the second order, $O(\epsilon^2)$, we obtain
\begin{align}
\label{surface_S_GH}
%& {\cal S}^{(2)}_\text{V}+{\cal S}^{(2)}_\text{GH}=\frac{1}{32\pi G_3\ell}\int dtdy
 { \delta^2 \mathcal{S} }_\text{V}
  + { \delta^2 \mathcal{S} }_\text{GH}
 & =\frac{ {\, \epsilon^2} }{16\pi G_3\ell}\int dtdy
\Biggl[f(TT'+YY'-3UU') \nonumber \\
&{} \qquad \qquad + \frac{1}{2u}\{3(1+f)U^2-(f-1)(T^2-Y^2)\}+\frac{1+f}{u}TY\Biggr]\Biggr{|}_{u\to 0}
\nonumber \\
%&=\frac{1}{32\pi G_3\ell}\int dtdy\left\{\frac{1+f}{u}TY-f(TY)'  \right\}, 
 & =\frac{{\, \epsilon^2} }{16\pi G_3\ell}\int dtdy\left\{\frac{1+f}{u}TY-f(TY)'  \right\} \Biggr{|}_{u\to 0} \,, 
\end{align} 
%where ${\cal S}^{(2)}_\text{V}:=\delta^{(2)}{\cal S}_\text{V}/(2\epsilon^2)$ and 
where % ${\cal S}^{(2)}_\text{V}:=\delta^{(2)}{\cal S}_\text{V}/(2\epsilon^2)$ and 
we have used (\ref{TT_condition2}) and $f'=(f-1)/u$ in the second equality. 
From eqs.~(\ref{TT_condition2}), (\ref{sol_global}), and (\ref{sol_BTZ}), we find that $U$, $T$, and $Y$ asymptotically behave as 
\begin{align}
\label{asymp_UTY}
& U\simeq C_1u^{\frac{3-p}{2}}+C_2u^{\frac{3+p}{2}} \,, \nonumber \\
& T\simeq -p(C_1u^{\frac{1-p}{2}}-C_2u^{\frac{1+p}{2}}) \,, \nonumber \\
& Y\simeq p(C_1u^{\frac{1-p}{2}}-C_2u^{\frac{1+p}{2}}) \,. % \nonumber \\
\end{align}
Substituting eqs.~(\ref{asymp_UTY}) into eqs.~(\ref{surface_S_GH}) and 
(\ref{ex_Sct_2}), one obtains
\begin{align}
\label{asmp_S_GH_ct}
%& {\cal S}^{(2)}_\text{V}+{\cal S}^{(2)}_\text{GH}=\frac{p^2}{32\pi G_3\ell}\int dtdy
{ \delta^2 \mathcal{S} }_\text{V}
+ { \delta^2 \mathcal{S} }_\text{GH}
& =\frac{ { \, \epsilon^2}\, p^2}{16 \pi G_3\ell}\int dtdy
\{-C_1^2(1+p)u^{-p}+2C_1C_2+O(u^p)\} \,,   \nonumber \\
%& {\cal S}^{(2)}_\text{ct}=\frac{\alpha p^2}{16\pi G_3\ell}\int dtdy \{C_1^2u^{-p}-2C_1C_2+O(u^p)\}. 
{ \delta^2 \mathcal{S} }_\text{ct}
 & =\frac{{\, \epsilon^2}\, \alpha p^2}{8\pi G_3\ell}\int dtdy \{C_1^2u^{-p}-2C_1C_2+O(u^p)\} \,. 
\end{align}
Thus, the total of the surface terms at $O(\epsilon^2)$ reduces to a finite term 
\begin{align}
\label{final_tot_surface}
%{\cal S}^{(2)}_\text{V}+{\cal S}^{(2)}_\text{GH}+{\cal S}^{(2)}_\text{ct}
%=-\frac{p^3}{16\pi G_3\ell}\int dtdy \,C_1C_2 
% { \delta^2 \mathcal{S}^{\text{OS}} } = 
{ \delta^2 \mathcal{S} }_\text{V}
+ { \delta^2 \mathcal{S} }_\text{GH}
+ { \delta^2 \mathcal{S} }_\text{ct}
=-\frac{ {\, \epsilon^2}\, p^3}{8\pi G_3\ell}\int dtdy \,C_1C_2 \,, 
\end{align}
if and only if one chooses the parameter $\alpha$ as
\begin{align}
\label{para_alpha}
\alpha=\frac{1+p}{2} \,. 
\end{align}
Note that $\alpha=1$ when the backreaction from the vacuum expectation value of the 
stress-energy tensor $\Exp{\calT_{\mu\nu}}$ is negligible, i.e., when $G_3 \to 0$ by eqs.~(\ref{dimensionless_relation}) 
and (\ref{dimensionless_para}). 
This is the case for the AdS/CFT correspondence in the $2$-dimensional non-dynamical boundary theory~\cite{Balasubramanian:1999re}. Note also that the coefficient $\alpha$ in 
the counter term ${\cal S}_\text{ct}$ in eqs.~(\ref{two_boundary_terms}) is not determined 
by the state of the boundary theory, but by the dimensionless parameter $\gamma_3$ of the theory 
via eq.~(\ref{dimensionless_relation}).

Summarizing the above results all together---in particular, the fact that the semiclassical Einstein equations~(\ref{Basic_Eq}) yields that the integrand of the first line of eq.~(\ref{second_variation_eff}) vanishes, 
we finally obtain the on-shell value $ { \delta^2 \mathcal{S}^{\text{OS}} }$ of the second order variation of the total effective action (\ref{second_variation_eff}) as the right-hand side of the total surface terms~(\ref{final_tot_surface}).

The deviation ${\Delta} F$ of the free energy of our static semiclassical solutions constructed in Sec.~\ref{sec:3} from that of the corresponding (either global AdS$_3$ or BTZ) background is related to the total effective action by 
\begin{align}
%\delta F=-\delta S/\int dt. 
{\Delta} F
= - {\Delta \mathcal{S}^{\text{OS}} }/\int dt \, = - \left( \mathcal{S}^{\text{OS}}
- \Bar{\mathcal{S}}^{\text{OS}} \right)
/\int dt \,. 
\end{align} 
At $O(\epsilon^2)$, $\Delta F$ is evaluated as 
\begin{align}
%\delta F=\frac{\epsilon^2p^3}{16\pi G_3\ell}\int dy \,C_1C_2<0
{\Delta} F {\, = - \frac{1}{2}\, \delta^2 \mathcal{S}^{\text{OS}}/\int dt
}
=\frac{\epsilon^2p^3}{16\pi G_3\ell}\int dy \,C_1C_2<0
\end{align}
by the inequality~(\ref{sign_C1C2}).
This means that the free energy of the semiclassical solution with $\Exp{\calT_{\mu\nu}}\neq 0$ is 
smaller than that of the corresponding (either the global AdS$_3$ or BTZ) background solution with $\Exp{\calT_{\mu\nu}}=0$. 
Therefore the semiclassical AdS$_3$ solution with vanishing source term is thermodynamically unstable. 

%%%%%%%%%%%%%%%%%%%%%%%%%%%
\section{Summary and discussions}\label{sec:5}
%%%%%%%%%%%%%%%%%%%%%%%%%%%
We have investigated thermodynamic instabilities of $3$-dimensional asymptotically AdS solutions 
to the holographic semiclassical Einstein equations by computing the free energies of the solutions. We have considered AdS$_3$ with AdS$_4$ bulk dual as our background solution to the holographic semiclassical Einstein equations with vanishing source term,  $\Exp{\calT_{\mu\nu}} = 0$. Then, by considering the tensor-type perturbations with respect to the AdS$_4$ bulk dual, we have analytically constructed static asymptotically AdS$_3$ solutions to the semiclassical Einstein equations with non-vanishing CFT source term, $\Exp{\calT_{\mu\nu}}\neq 0$. These new solutions can be regarded as semiclassical AdS$_3$ solution with ``quantum hair.'' We have constructed two such semiclassically hairy AdS$_3$ solutions: the one with respect to the static BTZ black hole background, which is the same as that found in~\cite{Ishibashi:2023luz}, and the other with respect to the global AdS$_3$ with no horizon. The free energies of these semiclassically hairy solutions have been evaluated by inspecting the on-shell effective action composed of both the $3$-dimensional Einstein-Hilbert action of the boundary conformal metric and the AdS$_4$ bulk action.
%, which gives the vacuum stress-energy tensor of the $3$-dimensional quantum CFT matter. 
We have shown that the free energy of the semiclassically hairy AdS$_3$ solution is smaller than that of the AdS$_3$ solution (with respect to either BTZ black hole chart or the global AdS$_3$ chart) when the universal parameter $\gamma_3$ in (\ref{dimensionless_para}) exceeds the critical value, i.e., $\gamma_3>1$.

The existence of such non-trivial AdS solutions with quantum hair reminds us of spontaneous symmetry breaking, in which a less symmetric solution appears from a highly symmetric one when one varies a control parameter of the theory. In our case, the parameter is the universal parameter $\gamma_3$ in (\ref{dimensionless_para}), and less symmetric solutions with ``quantum hair'' appears when $\gamma_3$ exceeds the critical value. As discussed in~\cite{Ishibashi:2023luz}, $\gamma_3$ is given by the ratio between the magnitude of the stress-energy tensor $\Exp{\calT_{\mu\nu}}$ composed of the vacuum fluctuations and that of the (classical) stress-energy tensor $\calT^\Lambda_{\mu\nu}$ 
composed of the $3$-dimensional cosmological constant. 
Then, the phase transition is triggered when the vacuum fluctuations overcome the magnitude 
of $\calT^\Lambda_{\mu\nu}$. 
If this effect is universal, one expects that such a spontaneous symmetry breaking should occur, 
regardless of whether the CFT is strongly coupled or not. It would be interesting to construct semiclassical solutions in the framework of a free CFT in curved spacetime.   

One may wonder if such a phase transition occurs for other spacetimes, such as asymptotically flat or de Sitter spacetimes. In asymptotically de Sitter spacetime, for example, 
one would obtain linearized semiclassical Einstein equations in asymptotically de Sitter spacetime, 
just like the master equation~(\ref{master_U}). The regularity condition on the black hole horizon or 
at the center determines the solutions uniquely, except the amplitude. So, one can expect that the linearized solution 
would be generically singular at the cosmological horizon, and therefore there are no static semiclassical solutions that become asymptotically de Sitter spacetimes. 
Similarly, one may also expect that asymptotically flat spacetime would not admit any static semiclassical solutions. 
It would be interesting to consider whether such a no go theorem in asymptotically flat or de Sitter spacetimes holds.  
  
There are other directions to extend the present work.  
For example, it would be interesting to compare the present result with the braneworld quantum BTZ black  hole~\cite{Emparan:2020znc} and its limit toward the conformal boundary of the AdS$_4$ bulk. 
It would also be interesting to explore whether the similar type of instabilities found in this paper and associated phase transitions can occur in the case of higher dimensional AdS spacetimes. For example, in $4$ or higher even-dimensional AdS spacetime, there is a trace anomaly, 
where the length scale in the highly symmetric phase would vary with the universal control 
parameter, $\gamma_4$. Furthermore, in higher dimensional AdS$_d$ black hole with dimension $d>3$, 
the black hole horizon radius $r_H$ would affect the phase transition as a new additional control parameter. 
%If such a spontaneous symmetry breaking is universal, it would be interesting to explore the 
%perturbation around the new less symmetric semiclassical solutions to confirm the existence 
%of the Nambu-Goldstone mode, or the massless mode, just like the chiral symmetry breaking. 

\begin{center}
{\bf Acknowledgments}
\end{center}
We wish to thank Roberto Emparan for useful discussions. We are grateful to the long term workshop YITP-T-23-01 held at YITP,
Kyoto University, where a part of this work was done. 
This work is supported in part by JSPS KAKENHI Grant No. 15K05092, 20K03938 (A.I.), 20K03975 (K.M.), 17K05427(T.O.), and also supported by MEXT KAKENHI Grant-in-Aid for Transformative Research Areas A Extreme Universe No.21H05186 (A.I. and K.M.) and 21H05182.  

%%%%%%%%%%%%%%%%%%%%%%%%%%%
\appendix
%%%%%%%%%%%%%%%%%%%%%%%%%%%

%%%%%%%%%%%%%%%%%%%%%%%
\section{Variation formulas}
%%%%%%%%%%%%%%%%%%%%%%%
Under the tensor-type perturbation~(\ref{TT_condition}), the first order variations are  
\begin{align}
\label{first_vari_R}
 \delta (\sqrt{-\tilde{G}})
 &=\frac{ {\epsilon} }{2} \sqrt{-\Bar{g}}\,\Bar{g}^{\mu\nu}h_{\mu\nu}=0 \,, 
 \nonumber \\ 
 \delta {R}[\tilde{G} ] 
 &= \epsilon \left( 
                       -h_{\mu\nu}\Bar{{R}}[{\tilde G}]^{\mu\nu}+\Bar{\tilde{\nabla}}^M\Bar{\tilde{\nabla}}^N h_{MN}
                       -\Bar{\tilde{\nabla}}^M\Bar{\tilde{\nabla}}_Mh 
                \right) 
\nonumber \\
&= \epsilon \left( 
                      h_{\mu\nu}\frac{2{\bdyg}^{\mu\nu}}{\ell^2}+\Bar{D}^\mu\Bar{D}^\nu h_{\mu\nu}
                      -\Bar{\tilde{\nabla}}^M\Bar{\tilde{\nabla}}_Mh 
                 \right) =0 \,, 
\nonumber \\
 \delta {R}[\tilde{G}]_{\mu\nu} 
  &= \epsilon \left(
                       -\frac{1}{2}\Bar{\tilde{\nabla}}^M\Bar{\tilde{\nabla}}_M h_{\mu\nu}-\frac{3}{\ell^2}h_{\mu\nu} 
                  \right) \,, 
\end{align}
where $\tilde{\nabla}_M$ denotes the covariant derivative operator compatible with $\tilde{G}_{MN}$.  
The second order variations are 
\begin{align}
\label{second_variations}
{ \delta^{2} } \left( \sqrt{-\tilde{g}} \right) 
  &= -\frac{ {\epsilon^2} }{2}
          \sqrt{-\Bar{g}}h_{\mu\nu}h^{\mu\nu} \,, 
\nonumber \\
\delta \left( \tilde{\nabla}^M\tilde{\nabla}^N {\epsilon}\, h_{MN} \right) 
  &={\epsilon^2}
          \sqrt{-\Bar{g}}
        \left[ -\frac{1}{2} \left( h^{\mu\nu}h'_{\mu\nu} \right)'
               -\Bar{D}_\mu \left( h^{\nu\alpha}\Bar{D}_\alpha {h_\nu}^\mu \right)
               -\frac{1}{4}\Bar{D}_\mu \Bar{D}^\mu \left( {h_\alpha}^\beta {h_\beta}^\alpha \right) 
        \right] \,, 
\nonumber \\
\delta \left( \tilde{\nabla}^M\tilde{\nabla}_M {\epsilon} h \right) 
 &= \epsilon^2 \left\{ 
                           - \left(h^{\alpha\beta}h_{\alpha\beta} \right)'' 
                           -\Bar{D}^2 \left( h^{\alpha\beta}h_{\alpha\beta} \right)
                  \right\} \,. 
\end{align}
Substituting these into the second variations of ${R}[\tilde{G} ]$, one obtains 
\begin{align}
\label{second_Ricci_scalar} 
&{ \delta^{2}}{R}[\tilde{G}]
= {\epsilon^2 \bigg\{} 
  -\frac{1}{\ell^2}h_{\mu\nu}h^{\mu\nu}+\frac{1}{2}h^{\mu\nu} \left( \Bar{D}^2h_{\mu\nu}+h''_{\mu\nu} \right)
  +\left(h^{\mu\nu}h_{\mu\nu} \right)'' \nonumber \\
&{} \qquad {} \qquad \qquad 
 +\frac{3}{4}\Bar{D}^2 \left( h^{\mu\nu}h_{\mu\nu} \right)
-\Bar{D}_\mu\Bar{D}_\nu \left(h^{\alpha\nu}{h_\alpha}^\mu \right)-\frac{1}{2}(h^{\mu\nu}h'_{\mu\nu})' { \bigg\} } \,.
\end{align}
Similarly, we also obtain the second variation of $\bdyR$ as 
\begin{align}
\label{second_bdyR}
& { \delta^{2} } \bdyR= { \epsilon^2\, \bigg\{ }
\frac{1}{2}h^{\mu\nu}\left(\Bar{D}^2-\frac{2}{\ell^2}\right)h_{\mu\nu}
+\frac{3}{4}\Bar{D}^2\left( h^{\mu\nu}h_{\mu\nu} \right)
-\Bar{D}_\mu\Bar{D}_\nu \left(h^{\alpha\nu}{h_\alpha}^\mu \right) { \bigg\} } \,. 
\end{align}

%%%%%%%%%%%%%%%%%%%%%%%%%%%%%%%%%%%%%%%%%%%%%%%%%%%%%%%%%%%%%%%%%%%%%%%%%%%%%%%%


\begin{thebibliography}{99}
%%%%%%%%%%%%%%%%%%%%%%%%%%%%%%%%%%%%%%%%%%%%%%%%%%%%%%%%%%%%%%%%%%%%%%%%%%%%%%%%
%\cite{Horowitz:1978fq}
\bibitem{Horowitz:1978fq}
G.~T.~Horowitz and R.~M.~Wald,
``Dynamics of Einstein's Equation Modified by a Higher Order Derivative Term,''
Phys. Rev. D \textbf{17} (1978), 414-416
%doi:10.1103/PhysRevD.17.414
%84 citations counted in INSPIRE as of 01 Nov 2023

%\cite{Horowitz:1980fj}
\bibitem{Horowitz:1980fj}
G.~T.~Horowitz,
``SEMICLASSICAL RELATIVITY: THE WEAK FIELD LIMIT,''
Phys. Rev. D \textbf{21}, 1445-1461 (1980)
%doi:10.1103/PhysRevD.21.1445
%108 citations counted in INSPIRE as of 22 Nov 2023


%\cite{Suen:1989bg}
\bibitem{Suen:1989bg}
W.~M.~Suen,
``Minkowski Space-time Is Unstable in Semiclassical Gravity,''
Phys. Rev. Lett. \textbf{62} (1989), 2217-2220
%doi:10.1103/PhysRevLett.62.2217
%29 citations counted in INSPIRE as of 01 Nov 2023

%\cite{Starobinsky:1980te}
\bibitem{Starobinsky:1980te}
A.~A.~Starobinsky,
``A New Type of Isotropic Cosmological Models Without Singularity,''
Phys. Lett. B \textbf{91} (1980), 99-102
%doi:10.1016/0370-2693(80)90670-X
%6457 citations counted in INSPIRE as of 01 Nov 2023

%\cite{Vilenkin:1985md}
\bibitem{Vilenkin:1985md}
A.~Vilenkin,
``Classical and Quantum Cosmology of the Starobinsky Inflationary Model,''
Phys. Rev. D \textbf{32} (1985), 2511
%doi:10.1103/PhysRevD.32.2511
%394 citations counted in INSPIRE as of 01 Nov 2023


\bibitem{Compere:2008us}
  G.~Compere and D.~Marolf,
  ``Setting the boundary free in AdS/CFT,''
  Class. Quant. Grav. \textbf{25}, 195014 (2008)
 doi:10.1088/0264-9381/25/19/195014
  [arXiv:0805.1902 [hep-th]].

\bibitem{Ishibashi:2023luz}
A.~Ishibashi, K.~Maeda and T.~Okamura,
``Semiclassical Einstein equations from holography and boundary dynamics,''
JHEP \textbf{05} (2023), 212
%doi:10.1007/JHEP05(2023)212
[arXiv:2301.12170 [hep-th]].
%4 citations counted in INSPIRE as of 02 Aug 2023

%\cite{Maldacena:1997re}
\bibitem{Maldacena:1997re}
J.~M.~Maldacena,
``The Large N limit of superconformal field theories and supergravity,''
Adv. Theor. Math. Phys. \textbf{2} (1998), 231-252
%doi:10.4310/ATMP.1998.v2.n2.a1
[arXiv:hep-th/9711200 [hep-th]].
%19044 citations counted in INSPIRE as of 01 Nov 2023

%\cite{Gubser:1998bc}
\bibitem{Gubser:1998bc}
S.~S.~Gubser, I.~R.~Klebanov and A.~M.~Polyakov,
``Gauge theory correlators from noncritical string theory,''
Phys. Lett. B \textbf{428} (1998), 105-114
%doi:10.1016/S0370-2693(98)00377-3
[arXiv:hep-th/9802109 [hep-th]].
%10165 citations counted in INSPIRE as of 01 Nov 2023

%\cite{Witten:1998qj}
\bibitem{Witten:1998qj}
E.~Witten,
``Anti-de Sitter space and holography,''
Adv. Theor. Math. Phys. \textbf{2} (1998), 253-291
%doi:10.4310/ATMP.1998.v2.n2.a2
[arXiv:hep-th/9802150 [hep-th]].
%12173 citations counted in INSPIRE as of 01 Nov 2023

%\cite{deHaro:2000vlm}
\bibitem{deHaro:2000vlm}
S.~de Haro, S.~N.~Solodukhin and K.~Skenderis,
``Holographic reconstruction of space-time and renormalization in the AdS / CFT correspondence,''
Commun. Math. Phys. \textbf{217}, 595-622 (2001)
%doi:10.1007/s002200100381
[arXiv:hep-th/0002230 [hep-th]].
%1608 citations counted in INSPIRE as of 11 Jan 2023

%\cite{Balasubramanian:1999re}
\bibitem{Balasubramanian:1999re}
V.~Balasubramanian and P.~Kraus,
``A Stress tensor for Anti-de Sitter gravity,''
Commun. Math. Phys. \textbf{208} (1999), 413-428
%doi:10.1007/s002200050764
[arXiv:hep-th/9902121 [hep-th]].
%1849 citations counted in INSPIRE as of 28 Oct 2023 
%

%\cite{Simon:1990jn}
\bibitem{Simon:1990jn}
J.~Z.~Simon,
``The Stability of flat space, semiclassical gravity, and higher derivatives,''
Phys. Rev. D \textbf{43}, 3308-3316 (1991)
%doi:10.1103/PhysRevD.43.3308
%115 citations counted in INSPIRE as of 26 Nov 2023

%\cite{Simon:1991bm}
\bibitem{Simon:1991bm}
J.~Z.~Simon,
``No Starobinsky inflation from selfconsistent semiclassical gravity,''
Phys. Rev. D \textbf{45}, 1953-1960 (1992)
%doi:10.1103/PhysRevD.45.1953
%72 citations counted in INSPIRE as of 26 Nov 2023

%\cite{Emparan:2020znc}
\bibitem{Emparan:2020znc}
R.~Emparan, A.~M.~Frassino and B.~Way,
``Quantum BTZ black hole,''
JHEP \textbf{11}, 137 (2020)
%doi:10.1007/JHEP11(2020)137
[arXiv:2007.15999 [hep-th]].
%50 citations counted in INSPIRE as of 28 Nov 2023

\end{thebibliography}
\end{document}